\documentclass[aip,jvst,reprint]{revtex4-1}
\usepackage{graphicx}
\usepackage{ulem}
\usepackage{amssymb}
\usepackage{amsmath}
\usepackage{wasysym}
\usepackage{relsize}
\begin{document}
%\draft                    

%\flushbottom
%\twocolumn[
%\hsize\textwidth\columnwidth\hsize\csname @twocolumnfalse\endcsname

\newcommand{\be}{\begin{equation}}
\newcommand{\ee}{\end{equation}}
\newcommand{\bea}{\begin{eqnarray}}
\newcommand{\eea}{\end{eqnarray}}
\newcommand{\Tbar}{{\bar{T}}}
\newcommand{\En}{{\cal E}}
\newcommand{\K}{{\cal K}}
\newcommand{\U}{{\cal U}}
\newcommand{\GC}{{\cal \tt G}}

\newcommand{\Lop}{{\cal L}}
\newcommand{\DB}[1]{\marginpar{\footnotesize DB: #1}}
\newcommand{\q}{\vec{q}}
\newcommand{\kt}{\tilde{k}}
\newcommand{\Lopn}{\tilde{\Lop}}
\newcommand{\noi}{\noindent}
\newcommand{\ovn}{\bar{n}}
\newcommand{\ovx}{\bar{x}}
\newcommand{\ovE}{\bar{E}}
\newcommand{\ovV}{\bar{V}}
\newcommand{\ovU}{\bar{U}}
\newcommand{\ovJ}{\bar{J}}
\newcommand{\calE}{{\cal E}}
\newcommand{\ovphi}{\bar{\phi}}
\newcommand{\zt}{\tilde{z}}
\newcommand{\ttl}{\tilde{\theta}}
\newcommand{\nuv}{\rm v}
\newcommand{\ds}{\Delta s}
\newcommand{\fn}{{\small {\rm  FN}}}
\newcommand{\cc}{{\cal C}}
\newcommand{\cd}{{\cal D}}
\newcommand{\tth}{\tilde{\theta}}
\newcommand{\cb}{{\cal B}}
\newcommand{\cg}{{\cal G}}
\newcommand\norm[1]{\left\lVert#1\right\rVert}

\title{ Electrostatic shielding versus anode-proximity effect in large area field emitters}
%\title{Minimizing electrostatic shielding with anode-proximity effect in large area field emitters}

%\vskip 0.3 in

\author{Debabrata Biswas}
\author{Rashbihari Rudra}
\affiliation{
Bhabha Atomic Research Centre,
Mumbai 400 085, INDIA}
\affiliation{Homi Bhabha National Institute, Mumbai 400 094, INDIA}

%\pacs{85.45.-w}{%\pacs{03.65.Sq}{}
%\pacs{03.65.Xp}{}
%\pacs{52.59.Sa}{}

\begin{abstract}
  Field emission of electrons crucially depends on the enhancement of the local electric field
  around nanotips. The enhancement is maximum when individual emitter-tips are well separated.
  As the distance between two or more nanotips decreases,
  the field  enhancement at individual tips reduces due to the shielding effect.  The anode-proximity effect
  acts in quite the opposite way, increasing the local field as the anode is brought closer to the emitter.
  For isolated emitters, this effect is pronounced when the anode is at a distance less than three times the height
  of the emitter. It is shown here that for a large area field emitter (LAFE),
  the anode proximity effect increases dramatically and can counterbalance
  shielding effects to a large extent. Also, it is significant
  even when the anode is far away. The apex field enhancement factor for a LAFE in the presence of an anode is
  derived using the line charge model. It is found to explain the observations well and can accurately predict the
  apex enhancement factors. The  results are supported by numerical studies using COMSOL Multiphysics.   
\end{abstract}

\maketitle

\section{Introduction}
\label{sec:intro}

Large area field emitters (LAFE) hold much promise as a high brightness source of cold electrons
\cite{spindt68,spindt91,teo,li2015}. The basic underlying
idea is the use of local electric field enhancement near the emitter apex \cite{edgcombe2001,forbes2003}
to lower the tunneling barrier at individual
nanotipped emitter sites, and, at the same time pack sufficient number of them to generate macroscopically
significant currents. There is a limit however on the mean separation between emitters since packing them more
densely can actually reduce the net current density of the LAFE due to shielding by
neighbouring emitters\cite{read_bowring,cole2014,zhbanov,harris15,forbes2016,db_rudra,rr_db_2019}.
This results in a reduced local field enhancement at emitter sites and hence a lowering of emission current.

While it is not possible to beat shielding altogether, the existence of a local field enhancing effect due to
the proximity of the anode \cite{wang2004,smith2005,podenok,pogo2009,pogo2010,jap16,lenk2018,db_anodeprox},
holds some promise in counter-balancing the former. The anode-proximity effect
has not been studied before from the LAFE point of view. For isolated emitters however, it is now well studied
numerically as well as analytically. It is known for instance that when the anode is close to the emitter,
there is a significant increase in local field at the emitter tip.
As the anode is moved further away, the effect reduces and practically ceases to exist when the anode
is separated from the cathode by about 3 times of height of the emitter. This distance is often set as
a thumb rule for the anode-at-infinity effect and it works quite well for an isolated emitter.

\begin{figure}[hbt]
  \begin{center}
%    \vskip 0.5cm
%\hskip -1.8cm
\hspace*{0.20cm}\includegraphics[scale=0.4,angle=0]{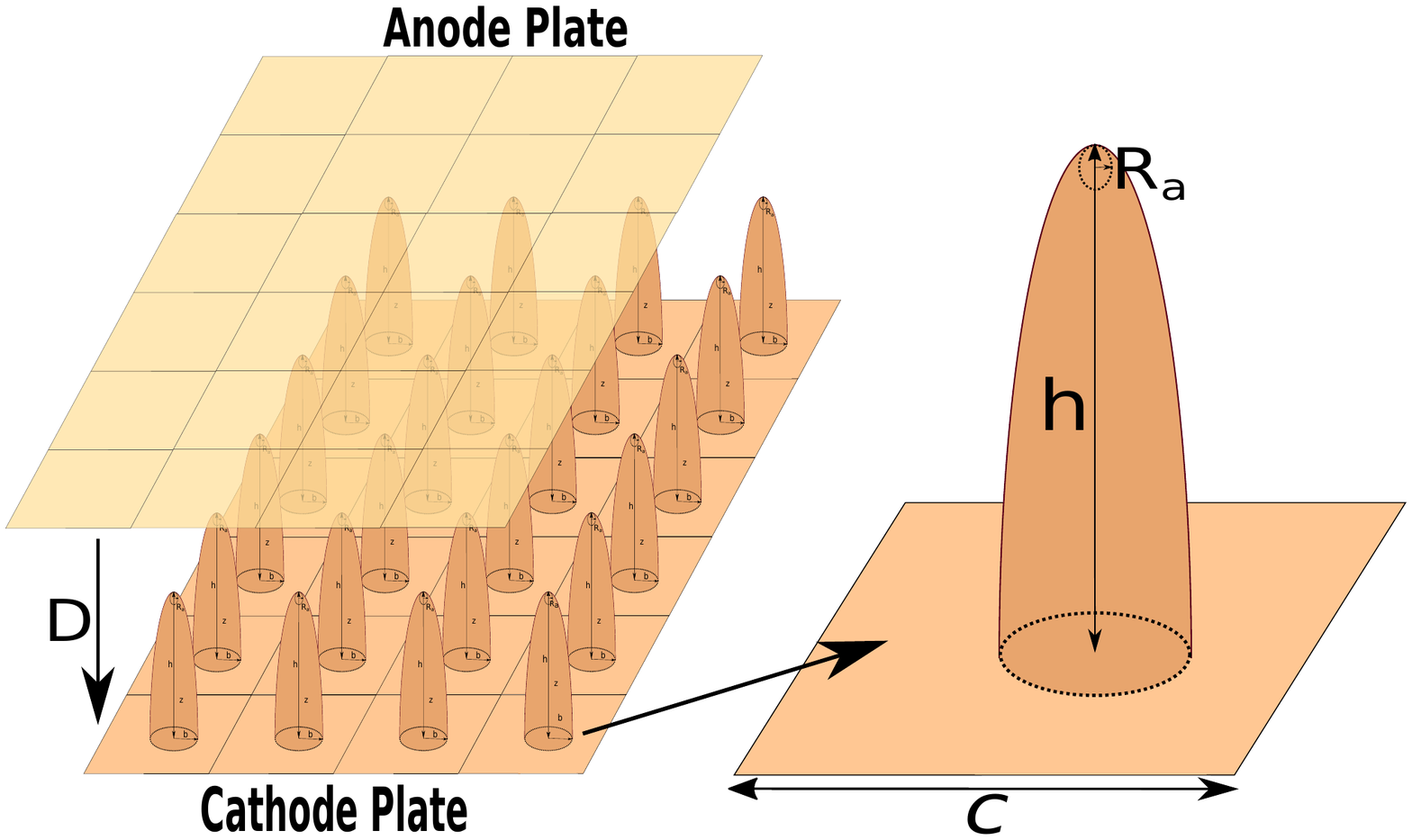}
\vskip -6.0 cm
\caption{Schematic of a large area field emitter array in a diode configuration. The
  distance between the anode and cathode plates is $D$ while individual emitters have a height
  $h$ and apex radius of curvature $R_a$. In general, individual emitters may be distributed randomly
  on the cathode plane. Typically, a finite sized LAFE has a few thousand emitters.}
\label{fig:diode_array}
\end{center}
\end{figure}

For a LAFE, the anode-proximity effect can in fact be enhanced further by bringing emitters closer. To see this, consider
a square lattice of nano-emitters (see Fig.~\ref{fig:diode_array}).
Each of them has an infinite number of images as a result of successive
reflections from the anode and cathode planes. As the lattice constant $c$ decreases, the number of images
within the zone of influence of a central emitter increases and starts contributing.
This leads to an enhanced anode-proximity effect and the local field increases substantially as
compared to the anode-at-infinity for the same lattice spacing. As an illustrative example \cite{using_COMSOL}, for an
emitter of height $h = 1500~\mu$m and apex radius of curvature $R_a = 1.5~\mu$m, the apex field
enhancement factor (AFEF) of an isolated, anode-at-infinity emitter is $\gamma_a(\infty,\infty) \simeq 317$. When
placed in a square array with lattice constant $c = h$, the enhancement factor is
$\gamma_a(D=\infty,c=h) \simeq 218$ with the anode still at `infinity'.
As the anode-cathode distance $D$ is reduced to $D = 1.5h$, the field enhancement in
a square array increases to $\gamma_a(D=1.5h,c=h) \simeq 303$ while
$\gamma_a(D=1.5h,c=\infty) \simeq 327$. Thus shielding dominates when the anode is at infinity
while anode-proximity has a dramatic effect when the emitters are packed closely, counter-balancing the
field-enhancement lowering effect of shielding. In this light, it need not be surprising
if anode-proximity dominates shielding for some value of $D$ and $c$. There are other important
ramifications of this finding. The counterbalancing act ensures that the optimal (mean) spacing for maximum
current density can now be lower (depending
on the closeness of the anode) than the `roughly 2 times emitter height rule' that applies
for the anode-at-infinity \cite{harris15,jap16}. This also implies that since more emitters can be packed, the current
density itself can rise significantly. These are some of the things that we shall investigate in
this paper.

The combined effects of anode-proximity and shielding phenomenon can be understood in terms of the
line charge model (LCM) which already provides a platform for shielding and anode-proximity effects individually. 
For simplicity, we shall restrict ourselves to ellipsoidal emitters for which the line charge density
is linear when the anode and shielding contributions are neglected. A limitation of the LCM is the distortion
in shape (the zero-potential contour)  when other emitters are in very close proximity or the anode is within
the a few radii of curvature of the central emitter apex. Nevertheless, the values of field enhancement
factor can be used profitably, with errors generally small for the emitters that do contribute to field
emission in a random LAFE, or if the lattice constant $c > h/2$ in a square array\cite{rr_db_2019}.

In the following, we shall model a general LAFE together with a planar anode using the linear line charge model
and derive a general formula for the field enhancement factor that accounts for both shielding and anode-proximity.
The accuracy of the formula is then tested using the finite element software COMSOL. We show that the 
percentage change in AFEF due to the presence of anode increases as the spacing
between the emitters decreases. The results point to a more optimistic outlook for the net emitted current from
an array of emitters or a random LAFE when the anode is in close proximity.

\section{Line charge model for a LAFE in the presence of anode}
\label{sec:LCM}

The potential at any point ($\rho,z$) due to an isolated line charge of extent $L$
placed at (0,0) perpendicular to a grounded conducting plane ($z = 0$) in the presence of an
electrostatic field $-E_0 \hat{z}$ can be expressed as \cite{pogo2009,harris15,jap16,db_fef}

\be
\begin{split}
V(\rho,z) = & \frac{1}{4\pi\epsilon_0}\Bigg[ \int_0^L \frac{\Lambda(s)}{\big[\rho^2 + (z - s)^2\big]^{1/2}} ds ~
  - \\
  &  \int_0^L \frac{\Lambda(s)}{\big[\rho^2 + (z + s)^2\big]^{1/2}} ds \Bigg] + E_0 z \\
= & \frac{1}{4\pi\epsilon_0} \int_{-L}^L \frac{\Lambda(s)}{\big[\rho^2 + (z - s)^2\big]^{1/2}} ds + E_0 z \label{eq:pot}
\end{split}
\ee

\noi
where  $\Lambda(s) = \lambda s$ in the line charge density. Note that the grounded conducting plane
is modeled by an image line-charge. The zero-potential contour
corresponds to the surface of the desired emitter shape so that
the parameters defining the line charge distribution
including its extent $L$, can, in principle be calculated by imposing the requirement that
the potential should vanish on the surface of the emitter.

As a next step, consider a collection of identical line charges each of extent $L$, placed randomly or in a regular
array\cite{db_rudra}. Denote the separation between the $i^{th}$ and $j^{th}$ line charge by $\rho_{ij}$.
For convenience, let the $i^{th}$ line charge be placed at the origin. The potential $V_S(\rho,z)$ (the subscript `$S$' for
`shielding' effect) can be expressed as

\be
\begin{split}
V_S(\rho,z)& =   E_0 z + \frac{1}{4\pi\epsilon_0}\Big[ \int_{-L}^L \frac{\lambda_i s}{\big[\rho^2 + (z - s)^2\big]^{1/2}} ds ~
  + \\
  &  \sum_{j \neq i} \int_{-L}^L \frac{\lambda_j s}{\big[(x - x_{ij})^2 + (y - y_{ij})^2  + (z - s)^2\big]^{1/2}} ds \Big]  \label{eq:pot}
\end{split}
\ee

\noi
where $\vec{\rho} = x\hat{i} + y\hat{j}$ and $\rho_{ij}^2 = x_{ij}^2 + y_{ij}^2$. For an infinite array,
the $\lambda_j$ are identical and its value can be determined
by demanding that the potential vanishes at the apex ($0,0,h$) of the $i^{th}$ emitter
where $h \simeq L + R_a/2$ where $R_a$ is the apex radius of the curvature of the emitters.

The next step is the introduction of the anode, separated from the cathode plane by a distance $D$.
These can be modeled by successive images of all line charge pairs (the line charge and its first image
on the cathode plane as incorporated in Eq.~\ref{eq:pot}) from the anode and cathode planes. The potential
$V_{SA}$ with both the `shielding' and `anode' terms can be expressed as \cite{db_rudra,db_anodeprox}

\be
\begin{split}
  V_{SA}&(\rho,z)  = E_0 z +
   \frac{1}{4\pi\epsilon_0} \int_{-L}^L ds \Bigg[ \frac{\lambda_i s}{\sqrt{\rho^2 + (z - s)^2}} ~-  \\
    & \sum_{n=1}^\infty  \frac{\lambda_i s}{\sqrt{\rho^2 + (2nD - z - s)^2}} + \frac{\lambda_i s}{\sqrt{\rho^2 + (2nD + z - s)^2}}  \\
     & + \sum_{j \neq i}  \Bigg\{ \frac{\lambda_j s}{\sqrt{(x - x_{ij})^2 + (y - y_{ij})^2  + (z - s)^2}}  \\       
    & -  \sum_{n=1}^\infty  \frac{\lambda_j s}{\sqrt{(x - x_{ij})^2 + (y - y_{ij})^2 + (2nD - z - s)^2}} \\
  &  + \frac{\lambda_j s}{\sqrt{(x - x_{ij})^2 + (y - y_{ij})^2 + (2nD + z - s)^2}} \Bigg\}  \Bigg]  \label{eq:potsum}
\end{split}
\ee

\noi
where the second and third terms under the integral are due to the images of the $i^{th}$ emitter, the fourth
term is due to shielding alone and the fifth and sixth have contributions from images of the $j^{th}$ emitters.

We are interested in determining the field enhancement at the tip (apex) of the $i^{th}$ emitter. This can be
achieved by differentiating Eq.~(\ref{eq:potsum}) with respect to $z$ and evaluating at
$x=0, y=0$ (or $\rho= 0$) and $z = h$.
As in Ref. [\onlinecite{db_rudra}], it can be shown that the dominant term is

\be
  \frac{\partial V}{\partial z} {|_{(\rho=0,z=h)}} \simeq - \frac{\lambda_i}{4\pi\epsilon_0}\Big[ \frac{2hL}{h^2 - L^2} \Big].
    \label{eq:Vz}
\ee

\noi
so that the field at the apex is known if $\lambda_i$ can be evaluated.

On setting $V_{SA}(\rho,z) = 0$ in Eq.~(\ref{eq:potsum}), an expression for $\lambda_i$ can be obtained.
Thus,

\be
\lambda_i = - \frac{4\pi\epsilon_0 E_0}{\ln[(h+L)/(h-L)] - 2L/h - \alpha_A +  \alpha_{S_i} - \alpha_{{SA}_i}} 
\ee

\noi
where

\be
\begin{split}
\alpha_A = & \sum_{n=1}^\infty \Bigg[ \frac{(2nD - h)}{h} \ln\Big(\frac{2nD - h + L}{2nD - h - L}\Big) \\
  & - \frac{(2nD + h)}{h} \ln\Big(\frac{2nD + h + L}{2nD + h - L}\Big) \Bigg], \label{eq:alpA}
\end{split}
\ee

\be
\begin{split}
  \alpha_{S_i} = & \sum_{j\neq i}^{N} \frac{\lambda_j}{\lambda_i} \Bigg[ \frac{1}{h}\sqrt{\rho_{ij}^2 + (h - L)^2} - \frac{1}{h}\sqrt{\rho_{ij}^2 + (h + L)^2} \\
    & + \ln\Bigg(\frac{\sqrt{\rho_{ij}^2 + (h + L)^2} + h + L}{\sqrt{\rho_{ij}^2 + (h - L)^2} + h - L}\Bigg) \Bigg], \label{eq:alpS}
\end{split}
\ee

\noi
and

\be
\begin{split}
  \alpha_{{SA}_i} = & \sum_{n=1}^\infty \sum_{j\neq i}^{N} \frac{\lambda_j}{\lambda_i} \Bigg[
    \frac{\cd_{mm}}{h} - \frac{\cd_{mp}}{h} - \frac{\cd_{pm}}{h} + \frac{\cd_{pp}}{h} \\
    & + \frac{2nD - h}{h} \ln\Big(\frac{\cd_{mp} + 2nD - h + L}{\cd_{mm} + 2nD - h - L}\Big) \\
    & - \frac{2nD + h}{h} \ln\Big(\frac{\cd_{pp} + 2nD + h + L}{\cd_{pm} + 2nD + h - L}\Big) \Bigg] \label{eq:alpSA}
\end{split}
\ee

\noi
where

\bea
\cd_{mm} & = & \sqrt{\rho_{ij}^2 + (2nD - h - L)^2} \nonumber \\
\cd_{mp} & = & \sqrt{\rho_{ij}^2 + (2nD - h + L)^2} \nonumber \\
\cd_{pm} & = & \sqrt{\rho_{ij}^2 + (2nD + h - L)^2} \nonumber \\
\cd_{pp} & = & \sqrt{\rho_{ij}^2 + (2nD + h + L)^2} \nonumber
\eea

The field enhancement factor at the apex of the $i^{th}$ emitter is thus

\be
\gamma_a \simeq \frac{2h/R_a}{\ln\big(4h/R_a\big) - 2 - \alpha_A +  \alpha_{S_i} - \alpha_{{SA}_i}} \label{eq:gamSA}
\ee

\noi
where we have used the relation $L = h - R_a/2$.

In general, for a random collection of emitters, $\lambda_j \neq \lambda_i$ even if all the
emitters are of equal height $h$. Following Ref. [\onlinecite{db_rudra}], we shall assume
$\lambda_j/\lambda_i \simeq 1$ since geometric effects are expected to dominate at least when the
emitters are not too close. For an infinite array, $\lambda_j/\lambda_i = 1$ and Eq.~(\ref{eq:gamSA}) is
easier to verify.

\section{Numerical Results}

The results presented in the previous section apply to any collection of emitters, whether finite
or infinite in number and irrespective of whether they are distributed in an array or randomly.
In view of the limitations of numerical results in handling a large number of emitters, we shall,
for purposes of comparison, confine ourselves in this section to an infinite array
which can be simulated using appropriate boundary conditions and nominal computational resources.

An infinite array can be used to understand how anode-proximity can counterbalance shielding
effects and also test the predictions of the line charge model derived in section \ref{sec:LCM}.
We shall assume the central ($i^{th}$) emitter to be placed at the origin. Other emitters ($j^{th}$ emitters)
have position vectors $\vec{\rho} = c(m_1 \hat{x} + m_2 \hat{y})$ in the $z = 0$ plane where $c$ is the
lattice constant. For the numerical results presented here, all emitters have a height $h=1500\mu$m and
apex radius of curvature $R_a = 1.5\mu$m. They are placed in an infinite square lattice with
lattice constant $c$.

Computationally (i.e. using COMSOL v5.4), an infinite square array with lattice constant $c$ can be simulated by
imposing `zero surface charge density' at $x,y = \pm c/2$.
Thus, $\partial V/\partial (x,y) = 0$ at $x = \pm c/2$ and $y = \pm c/2$. The boundary condition
at the cathode and anode is Dirichlet with the cathode potential $V_C = 0$ and the anode potential
$V_A = D E_0$, where $D$ is the anode-cathode distance and $E_0$ is the macroscopic field.
In all the calculations presented here, a `general physics' mesh type is used with
the minimum element size smaller than $2\times 10^{-5}\mu$m, the
maximum element size smaller than $0.1\mu$m, the curvature factor smaller than 0.05
and the element growth rate 1.3. Convergence with respect to these parameters have been ensured \cite{agnol}. 

\begin{figure}[hbt]
  \begin{center}
    \vskip -0.5cm
%\hskip -1.8cm
\hspace*{-1.0cm}\includegraphics[scale=0.35,angle=0]{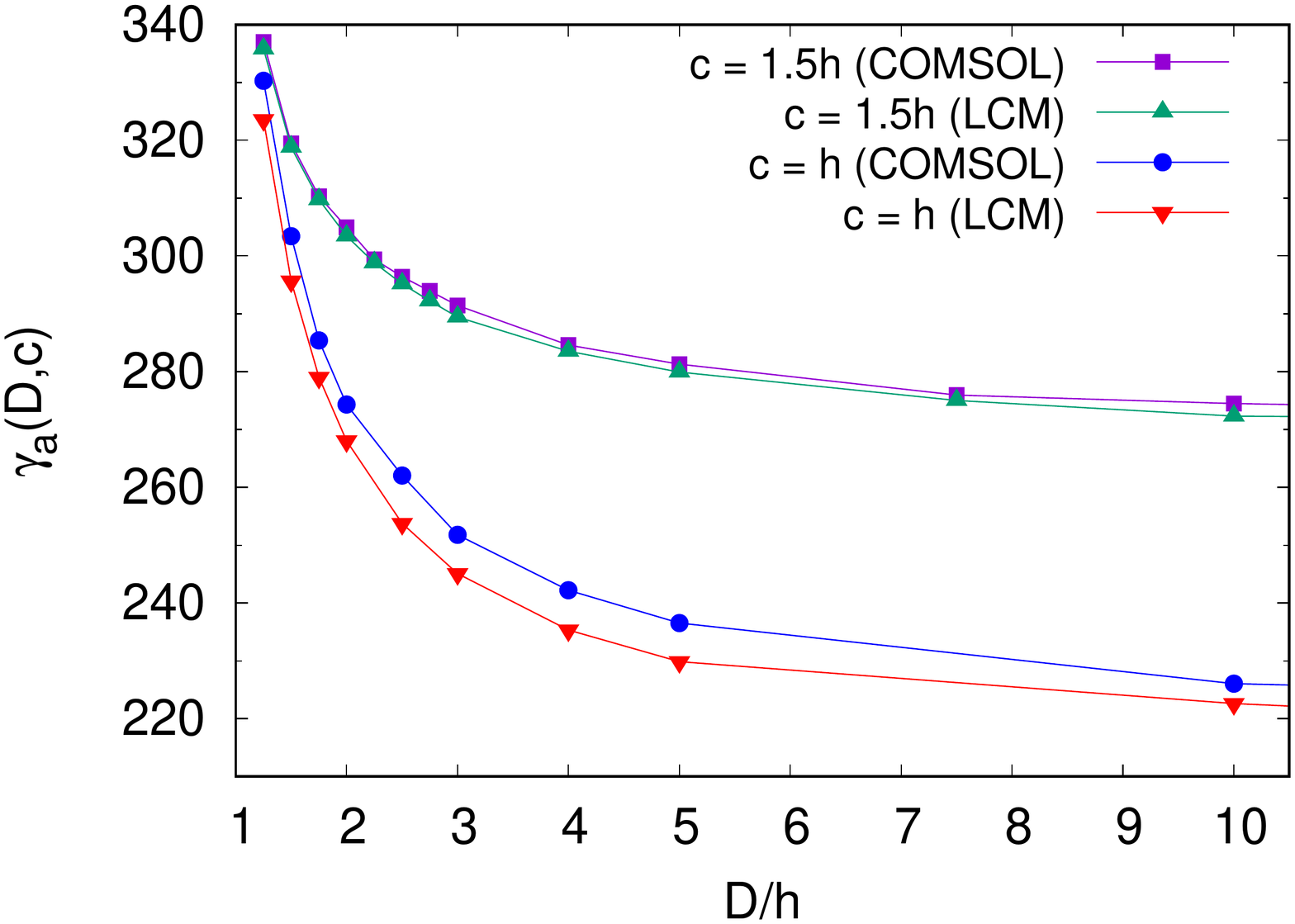}
\vskip -0.5 cm
\caption{The field enhancement factor $\gamma_a(D,c)$ as a function of anode-cathode distance $D$. Two values of
  lattice constant $c$ are considered and each value of $D$ and $c$, $\gamma_a(D,c)$ is evaluated using COMSOL (solid square for $c=h$
  and solid circle for $c = 1.5h$) and the LCM predictions of Eq.~(\ref{eq:gamSA}) (denoted by solid triangles). }
\label{fig:gamma_anode}
\end{center}
\end{figure}

\begin{figure}[hbt]
  \begin{center}
    \vskip -1.0cm
%\hskip -1.8cm
    \hspace*{-1.0cm}\includegraphics[scale=0.34,angle=0]{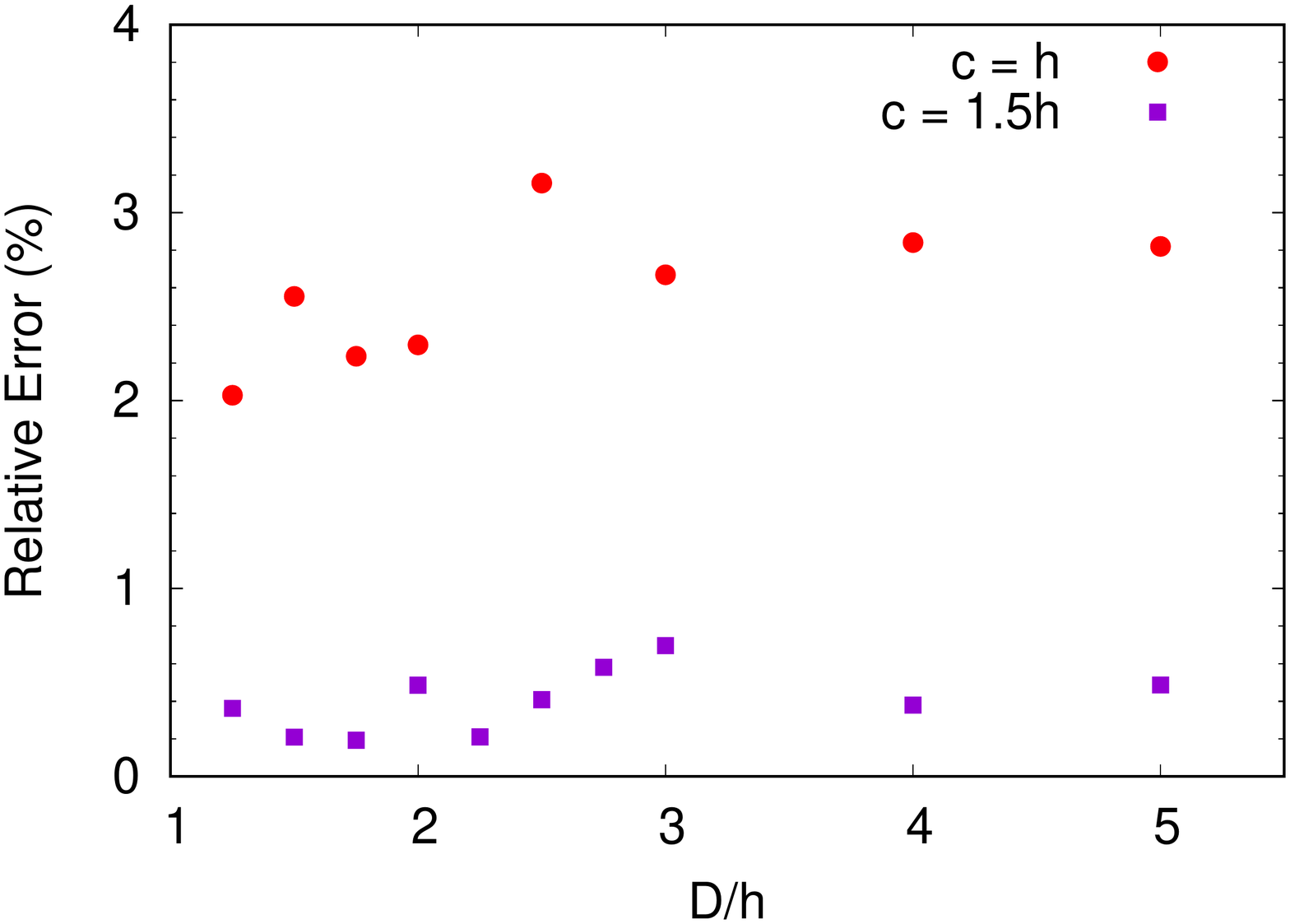}
\vskip -0.75 cm
\caption{The relative error in LCM prediction  for the two different lattice constant $c$
for various anode-cathode distance $D$.}
\label{fig:error}
\end{center}
\end{figure}

The apex field enhancement factor can be determined with Eqns~(\ref{eq:gamSA})
and (\ref{eq:alpA})-(\ref{eq:alpSA}) for each value of
$D$ and $c$. All emitters within a radius of $80c$ have been included while the number of images considered is
typically around 1000. The apex field enhancement has also been computed using the finite element software COMSOL.
The results are shown in Fig.~\ref{fig:gamma_anode}.

Clearly, at the smaller nearest-neighbour pin spacings, the anode-proximity effect is much
stronger \cite{using_COMSOL}. To see this,
note that $\gamma_a(D=1.25h,c=h) \simeq 330$ while $\gamma_a(D=10h,c=h) \simeq 226$ as compared
to $\gamma_a(D=1.25h,c=1.5h) \simeq 337$ and $\gamma_a(D=10h,c=1.5h) \simeq 275$. Thus, as compared to $D=10h$,
$\gamma_a$ increases by about $46\%$ at $D = 1.25h$ for $c = h$, while over the same
range, $\gamma_a$ increases by $22.5\%$ for $c=1.5h$. The effect is even
more dramatic for $c = 0.75h$ where $\gamma_a$ increases by $70\%$ with $\gamma_a(D=10h,c=0.75h) \simeq 185$ and
$\gamma_a(D=1.25h,c=0.75h) \simeq 315.3$.

We next study the predictions of line charge model. It is obvious from Figs.~\ref{fig:gamma_anode}
that the error is much smaller at larger nearest-neighbour pin spacings. This is quantified in
Fig.~\ref{fig:error} where the relative error defined as

\be
\text{Relative Error} (\%) = \frac{\gamma_a^{comsol}(D,c) -\gamma_a^{LCM}(D,c)}{\gamma_a^{comsol}(D,c)} \times 100
\ee
  
\noi
is plotted. Here $\gamma_a^{comsol}(D,c)$ and $\gamma_a^{LCM}(D,c)$ are respectively the values of the
apex field enhancement determined using COMSOL and LCM. For $c = 1.5h$, the average error in
prediction in the range $D = 1.25h$ to $D = 10h$
is about $0.43\%$ while for $c = h$, the average error increases to about $2.45\%$ in the same range of $D$.
At $c = 0.75h$, the average error
increases to about $7\%$ for $D \in [1.25h,10h]$ with errors about $5\%$ for $D \in  [1.25h,2h]$.

Thus, the line charge model captures the anode-proximity for arrays of emitters with errors that are generally below $2.5\%$
for $c \geq h$. It can thus be used to calculate the optimal spacing in the new light of enhanced anode-proximity
effect for arrays of emitters. Recall that when the anode is at infinity \cite{db_rudra}, the array current density is maximum
when the lattice constant (or mean spacing) is about $2h$, with a slight variation depending on the electric field. In the
presence of the anode, this optimal spacing is expected to change depending on the anode-cathode distance $D$.

\begin{figure}[hbt]
  \begin{center}
    \vskip -0.85cm
%\hskip -1.8cm
    \hspace*{-1.0cm}\includegraphics[scale=0.34,angle=0]{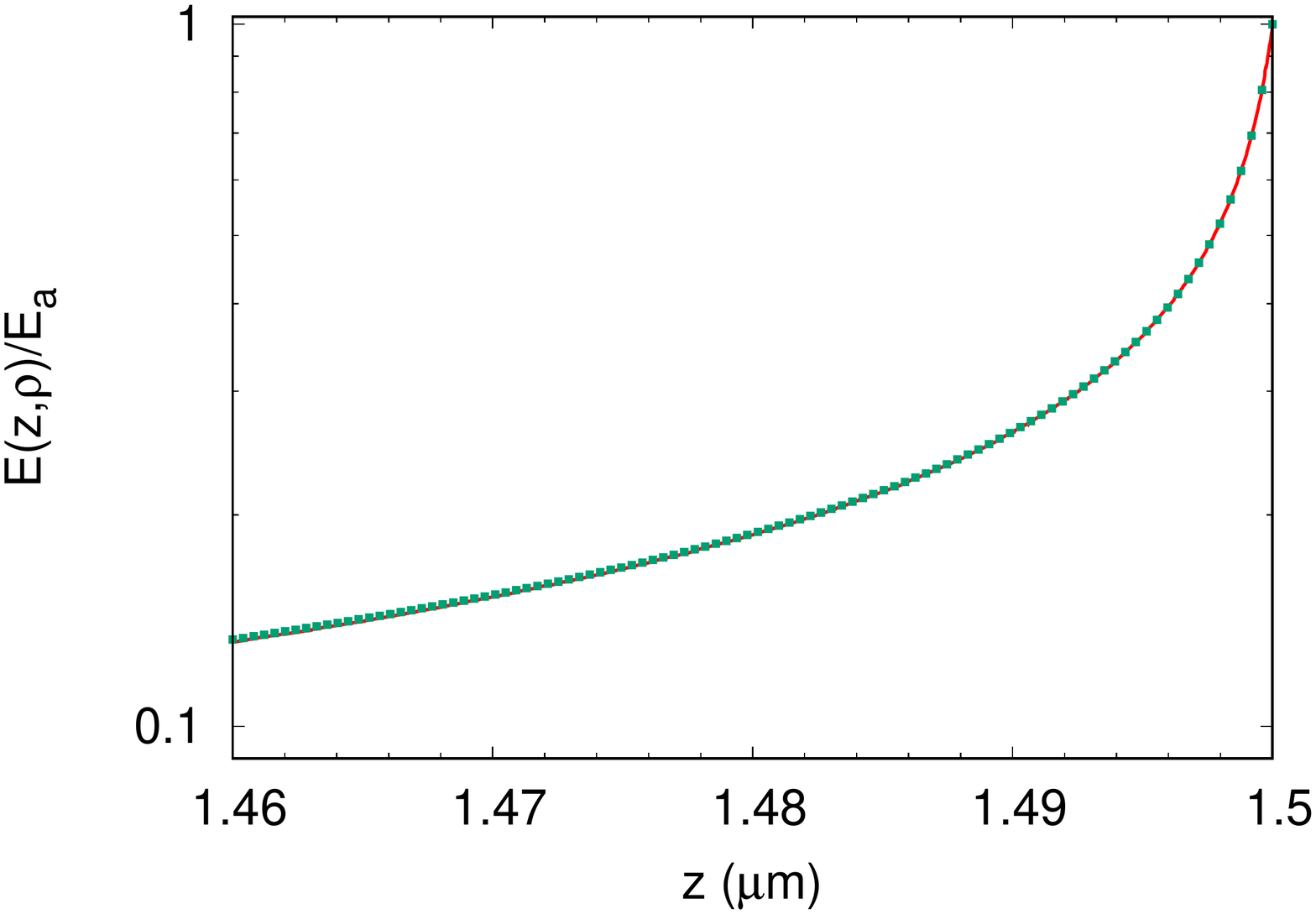}
\vskip -0.75 cm
\caption{Variation of the electric field on the emitter surface around the apex follows the generalized cosine
  law  $E(z,\rho)/E_a = \cos\tth = (z/h)/\sqrt{(z/h)^2 + (\rho/R_a)^2}$. The electric field data is obtained
  using COMSOL (solid curve) while the solid squares are obtained using the expression for $\cos\tth$.
Here $D = 1.25h$ and $c = h$.}
\label{fig:cosinelaw}
\end{center}
\end{figure}

For an infinite square lattice, the array current density $J_{array}$ is evaluated by calculating the current from a single
emitter pin and dividing by $c^2$. Thus, $J_{array} = I_{pin}/c^2$ where\cite{db_distrib}

\bea
I_{pin} & \simeq & 2\pi R_a^2 \cg J(0,h) \\
J(0,h) & =  & \frac{1}{(\mathlarger{t_F})^2} \frac{A_\fn}{W} E_a^2 \exp(-B_\fn v_{{\small F}} W^{3/2}/E_a) \\
\eea

\noi
where $\cg  = 1/(\cb D_1)$,  $f_0   \simeq   1.439965 E_a/W^2$, $D_1 = 1 - f_0/6$,
$v_{\small F}  =  1 - f_0 + (1/6) f_0 \ln f_0$, $t_{\small F}  =  1 + f_0/9 - (1/18) f_0 \ln f_0$
and $E_a = \gamma_a(D,c) E_0$. In the above $W$ is the work-function ({\rm eV}) and
$A_\fn~\simeq~1.541434~{\rm \mu A~eV~V}^{-2}$ and $B_\fn~\simeq~6.830890~{\rm eV}^{-3/2}~{\rm V~nm}^{-1}$
are the conventional Fowler-Nordheim constants\cite{FN,Nordheim,murphy,forbes,jensen_ency}.
The expression for $\cg$ is obtained using the generalized cosine
law\cite{db_ultram,cosine} of field variation near the apex, a result found to be true in the present scenario as shown in Fig.~\ref{fig:cosinelaw}.

\begin{figure}[hbt]
  \begin{center}
    \vskip -0.5cm
%\hskip -1.8cm
    \hspace*{-1.0cm}\includegraphics[scale=0.34,angle=0]{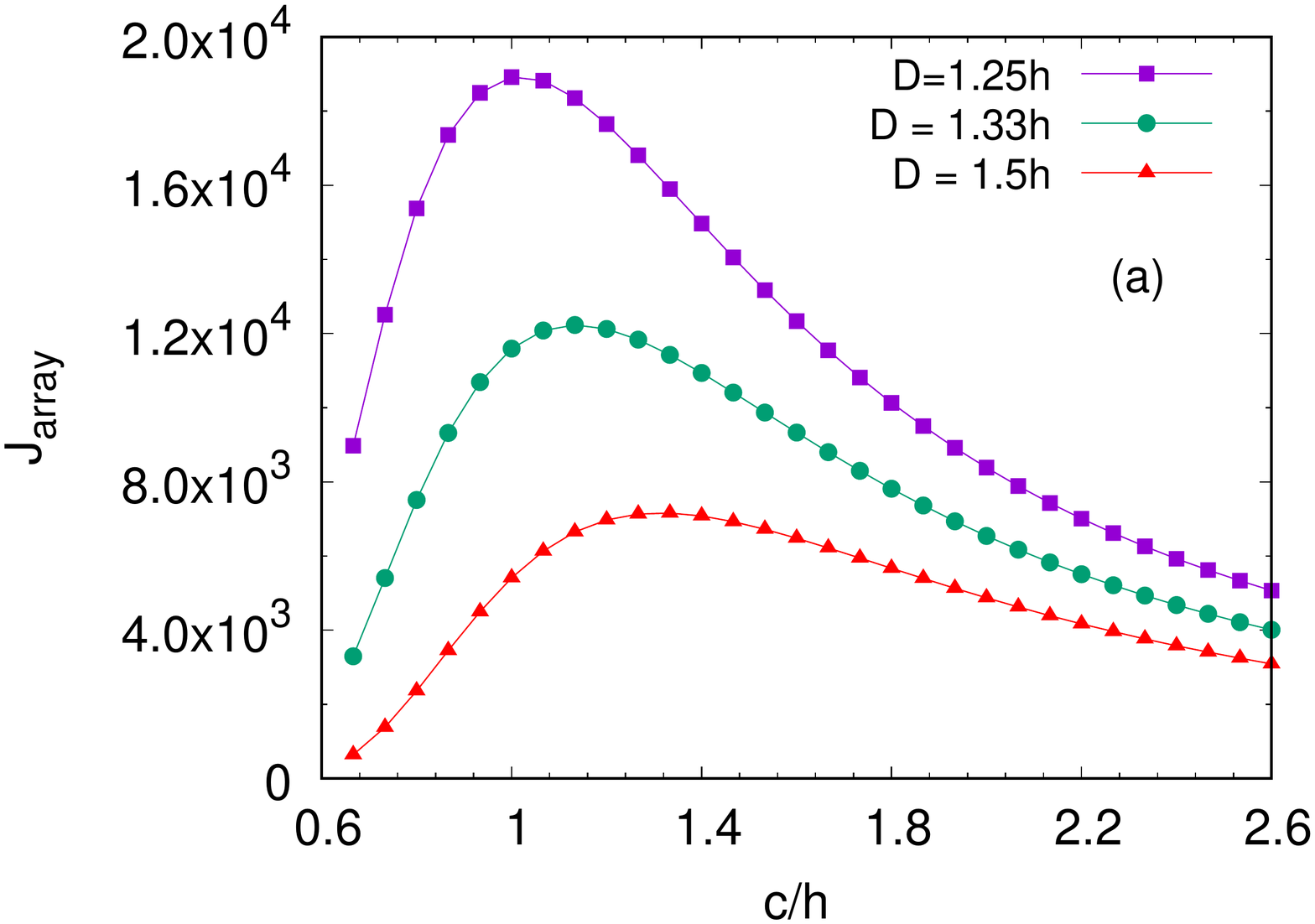}
    \hspace*{-1.0cm}\includegraphics[scale=0.34,angle=0]{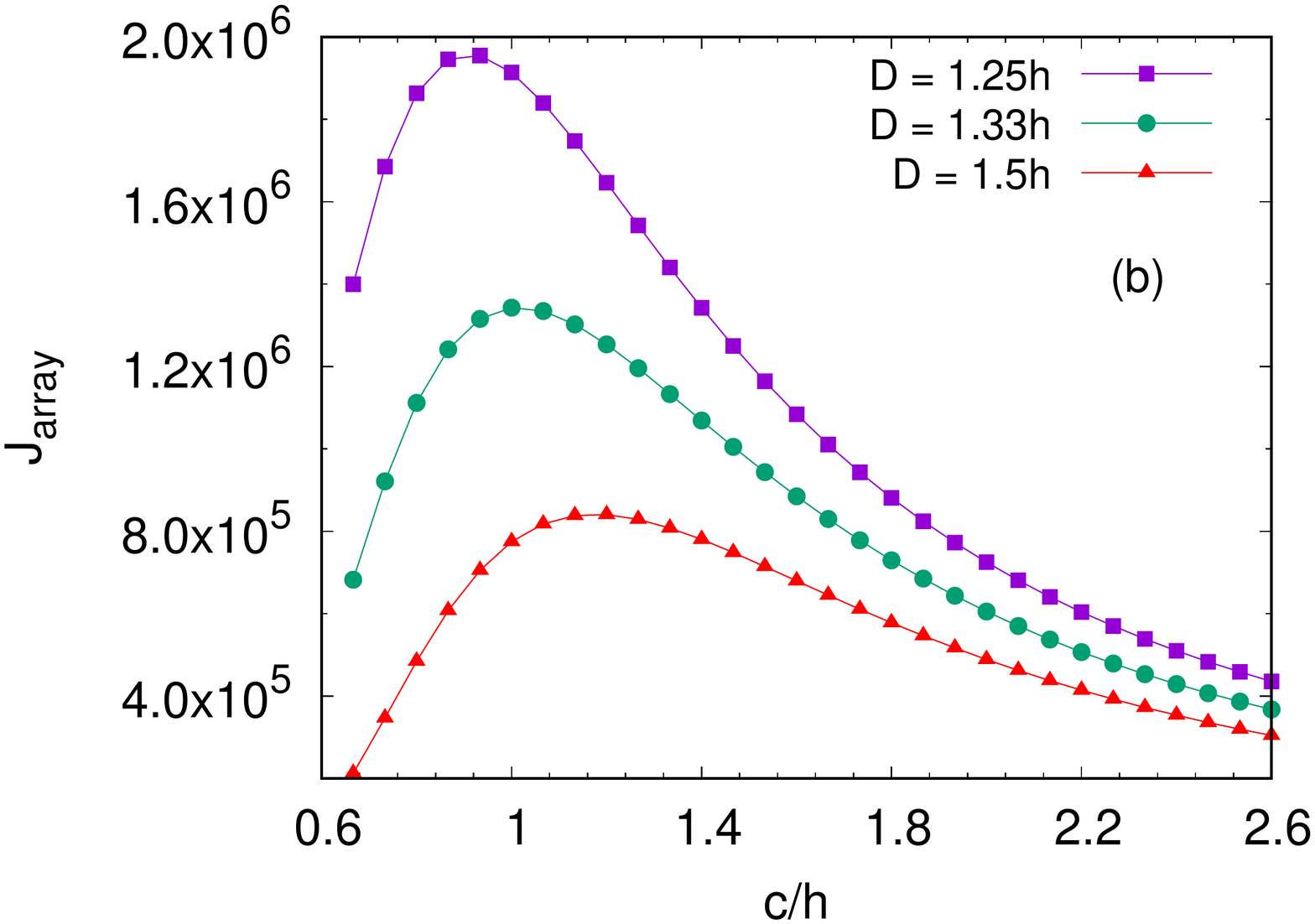}
\vskip -0.75 cm
\caption{Variation of the array current density $J_{array}~({\rm A/m^2})$ with lattice spacing $c$ for 3 different
  anode-cathode spacing $D$ and (a) $E_0 = 20$MV/m and (b) $E_0 = 30$MV/m. Note that the optimal spacing
shifts below the emitter height for $D = 1.25h$ and $E_0 = 30$MV/m.}
\label{fig:currentdensity}
\end{center}
\end{figure}

Fig.~\ref{fig:currentdensity} shows the array current density for 2 values of macroscopic electric field and
3 values of anode-cathode distance $D$. Clearly, as the anode comes closer to the emitter ($D$ decreases),
the optimal spacing becomes smaller and
the maximum current density itself increases at any given macroscopic field as evident from the figures. Further, at a
higher macroscopic field, the shift to smaller optimal spacing is greater. At $E_0 = 30$MV/m for instance, the
optimal spacing is $c \simeq 0.75h$, which is smaller than the height of the emitter. The significance of the
anode-proximity effect in a LAFE can be judged by noting that when the anode is far away ($D = 100h$), the maximal current density
is  $1.19 \times 10^3 \text{A/m}^2$ at $E_0 = 20$MV/m ($c/h = 2.4$) while at $E_0 = 30$MV/m, the maximal current
density is  $1.65\times 10^3$ ($c/h = 2.13$). 

\section{Discussions and Conclusions}

It is clear from the preceding analysis that for a LAFE, the anode-proximity effect plays a dominant role and
can counterbalance electrostatic shielding. We have also established that the line charge model
predicts the apex field enhancement factor accurately for lattice constants $c \geq 1.5h$ while
the error was found to be less than $5\%$ for $c = 0.75h$ and anode distance $D \leq 2h$.

The line charge model in fact under-predicts the apex field enhancement factor $\gamma_a(D,c)$ for small
values of $D$ and $c$. This is due to shape broadening of the zero-potential contour. The current densities
obtained using LCM thus provide a lower bound.

In the present study, the dimensions of the emitter chosen are such that curvature-corrections to the tunneling potential are
negligible (typically when $R_a > 100$nm, see [\onlinecite{db_curvature}]). A similar analysis can
be done for emitters with smaller apex radius of curvature (a few nanometers)
using the curvature-corrected formula for emission current\cite{db_curvature,db_gated}. Due to the nature of
the corrections \cite{db_curvature}, it is
easy to see that the current densities will decrease somewhat due to a slightly broadened tunneling barrier.
Nevertheless, anode-proximity will play a dominant role and enhance the optimal current density.
Note that though the emission current characteristics depend on the scale of the problem
(e.g. nanometer vs micrometer), Eq.~(\ref{eq:gamSA}) is independent of the scale  since the
quantities involved are ratios of lengths and hence are dimensionless.

Finally, even though we have chosen an ellipsoid to demonstrate the enhanced anode-proximity effect in a LAFE,
differently shaped emitters will also show the enhanced effect. The line charge model will also continue
to hold but must be replaced by a nonlinear line-charge, making predictions slightly more involved \cite{db_anodeprox}.
The way forward seems to be semi-analytical model for non-ellipsoidal shapes where the prefactor of the logarithmic
term in Eq.~(\ref{eq:gamSA}) and the ellipsoidal constant 2 are modified using numerical inputs such as in
[\onlinecite{db_anodeprox}] while $\alpha_A$, $\alpha_{SA}$ and $\alpha_S$ are calculated
analytically.

It is hoped that the results presented here will be useful in analysing LAFE emitter experiments
or in designing cathodes. Of particular interest will be finite-sized array
and a random distribution of emitters, both of which are presently beyond
the reach of purely numerical methods. The finite array cannot be reliably analysed using finite element
codes, especially ones  that are small enough for surface effects to be important and yet too large for simulation. 
A random distribution consisting of few thousand emitters is another situation where the method described here
can be used profitably. In both situations, the use of a gated-anode or a grid-anode in close proximity to the
emitter-tips will lead to enhancement of the local field, compared to an anode that is far away.
Experimental results on random LAFE such as in [\onlinecite{bieker2018}] where the emitter shape is
conical and $D \simeq 2h$, offer a scope for validation of the model presented here.  

\section{Acknowledgement}

The authors wish to thank Gaurav Singh for valuable help and discussions.

%\vskip -.75 in
$\;$\\
\section{References} 

%\begin{references}


\begin{thebibliography}{99}
\bibitem{using_COMSOL} The enhancement factors quoted here are obtained using COMSOL v5.4.
\end{thebibliography}
\end{document}